\documentstyle[11pt,moriond,epsfig]{article}

\def\Journal#1#2#3#4{{#1} {\bf #2}, #3 (#4)}

\def\NPB{{\em Nucl. Phys.} B}
\def\PLB{{\em Phys. Lett.}  B}

\def\PRD{{\em Phys. Rev.} D}
\def\ZPC{{\em Z. Phys.} C}

\def\be{\begin{equation}}
\def\ee{\end{equation}}
\def\bea{\begin{eqnarray}}
\def\eea{\end{eqnarray}}

\begin{document}
\vspace*{4cm}
\title{VECTOR MESON PRODUCTION AT HERA}

\author{ JAN  FIGIEL }

\address{ 
Institute of Nuclear Physics Cracow, Poland \\
e-mail:  Jan.Figiel@ifj.edu.pl\\
on behalf of the ZEUS and H1 collaborations}

\maketitle\abstracts{
New results on elastic electroproduction and proton-dissociative
photoproduction at large $|t|$ of $\rho, \phi$ and $J/\psi $ mesons are 
presented. They are interpreted within perturbative QCD. 
 }

\section{Introduction}

\begin{figure}[h] 
\begin{center} 
\epsfig{file=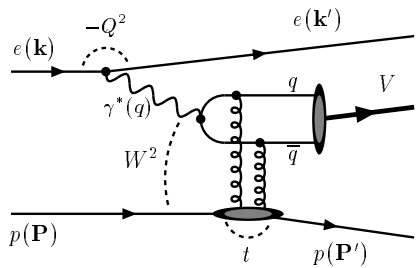,height=35mm}
\end{center}
\caption{The pQCD diagram for VM electro-production via 2-gluon exchange.
$W$ is energy in ~$\gamma^* p $ ~CMS, ~$Q^2$ ~is photon virtuality and ~$t$
~denotes square of the 4-momentum transfer between incoming and outgoing
proton. 
\label{fig:pqcddiag}}
\end{figure}

At high energy the diffractive electroproduction of the vector mesons (VM)
is a two-stage process: at first the virtual photon fluctuates into VM which
then interacts with the target. The latter is the process of interest and in 
analogy with soft hadron-hadron interactions can be interpreted 
within Regge phenomenology \cite{do}. However at large $Q^2$ or mass of the
meson $M_V$ (and in principle $|t|$) the range of the VM-p interaction is
small: it is ``hard'' and perturbative QCD can be applied~\cite{ry}.
It describes the process via  exchange of colour singlet system of gluons (2
gluons in leading order - Figure~\ref{fig:pqcddiag}). Contrary to Regge
approach, in pQCD framework  
the steep rise of the VM production cross section and violation of the
s-channel helicity conservation (SCHC) is predicted. The transition between
soft and hard regime  is a test ground of the QCD and is studied intensively at
HERA .

\section{Elastic electroproduction of $ \rho $ and $ J/\psi $ mesons}

\begin{figure}[t]
\vspace*{-1.9cm}
\begin{center}
\mbox{}
\psfig{file=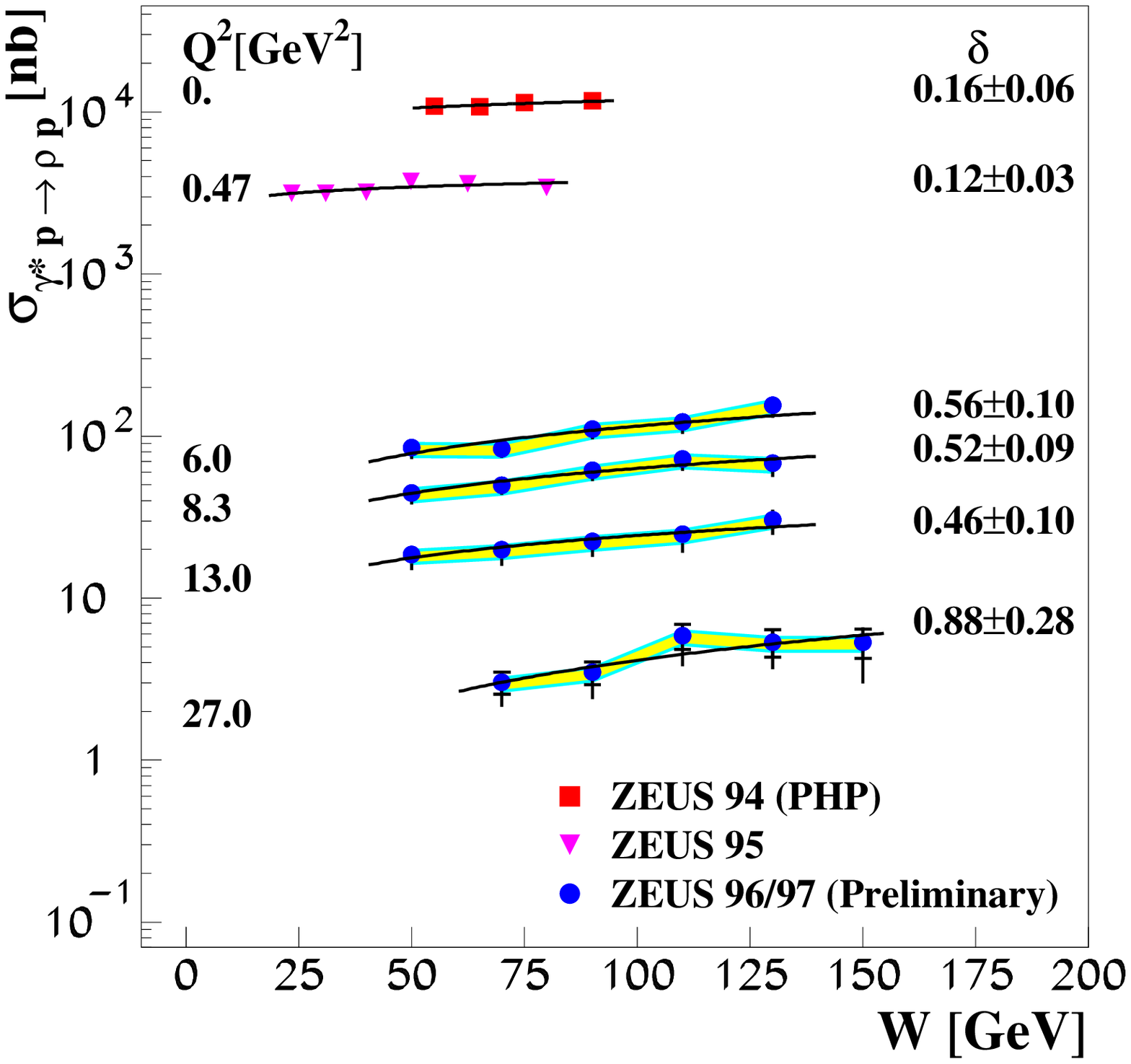,bbllx=30pt,bblly=170pt,bburx=510pt,bbury=620pt,height=6.6cm}
~~\epsfig{file=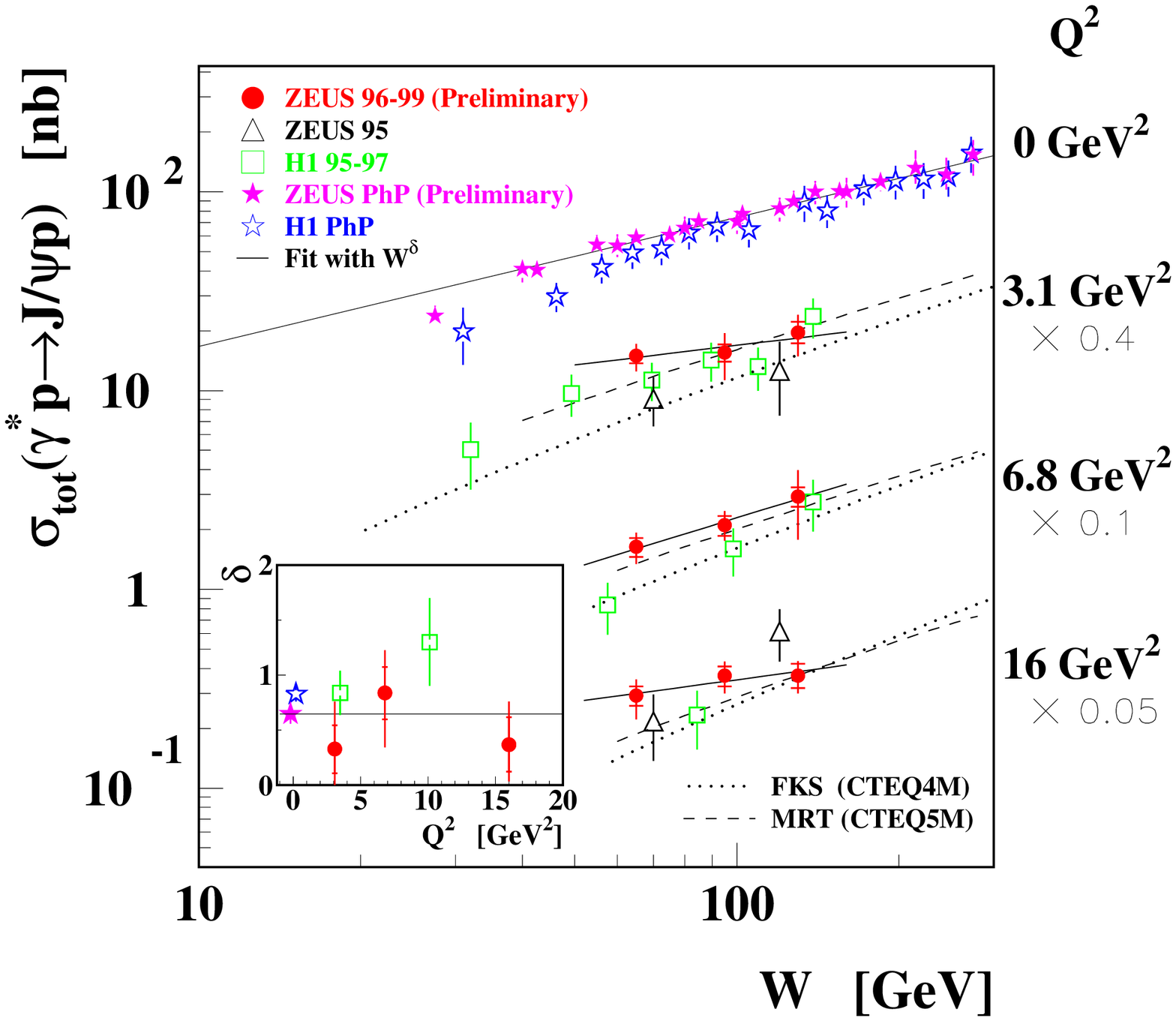,height=8.2cm}
\caption{Left - the $\rho $ cross section, ~right - the $J/\psi $ cross
section; solid lines represent fits to the ZEUS data with the formula $ \sim
W^{\delta} $ and dotted and dashed ones - the predictions of pQCD models (see
text). The inset displays the fitted value of $\delta$ as a function of $Q^2$.
\label{fig:rhopsi}}
\end{center} 
\end{figure}

Diffractive elastic electroproduction of $ \rho $ mesons has been measured by
the ZEUS collaboration~\cite{ro}. The cross section $ \sigma_{\gamma^*p
\rightarrow \rho p} $ is shown in Figure~\ref{fig:rhopsi} (left) as a function
of $W$ in different $Q^2$ intervals. It  is parametrised as $ \sim W^{\delta}
$; magnitude of power $\delta$ reflects steepness of its energy dependence.
At $Q^2 \sim 0 ~~\delta$ is close to 0.2, value typical
for soft hadronic interactions whereas at large $Q^2 $  it is much bigger
indicating strong energy dependence typical for hard colour singlet exchange.
Thus large $Q^2$ provides ``hard'' QCD scale.
On the right side of Figure~\ref{fig:rhopsi} ~the recent measurement of 
the $ J/\psi $ cross
section by ZEUS~\cite{psiz} and H1~\cite{psih} experiments are shown. As before
the energy dependence in $Q^2$ intervals is fitted with power law formula.
However now one observes steep rise of the cross section with $W$ ($\delta
\simeq 0.65 $) irrespectively of  $Q^2$  value: ``large'' $M_{\psi}^2$ seems
to be sufficient hard scale. The pQCD based models~\cite{fra}, relating the 
$J/\psi $ cross section to  the gluon density in proton reproduce the
behaviour of the data.

\section{Photoproduction of $ \rho $, $\phi$ and $ J/\psi $ mesons at large
$|t|$ }

\begin{figure}[t] 
\vspace*{-1.0cm}
\begin{center} 
\epsfig{file=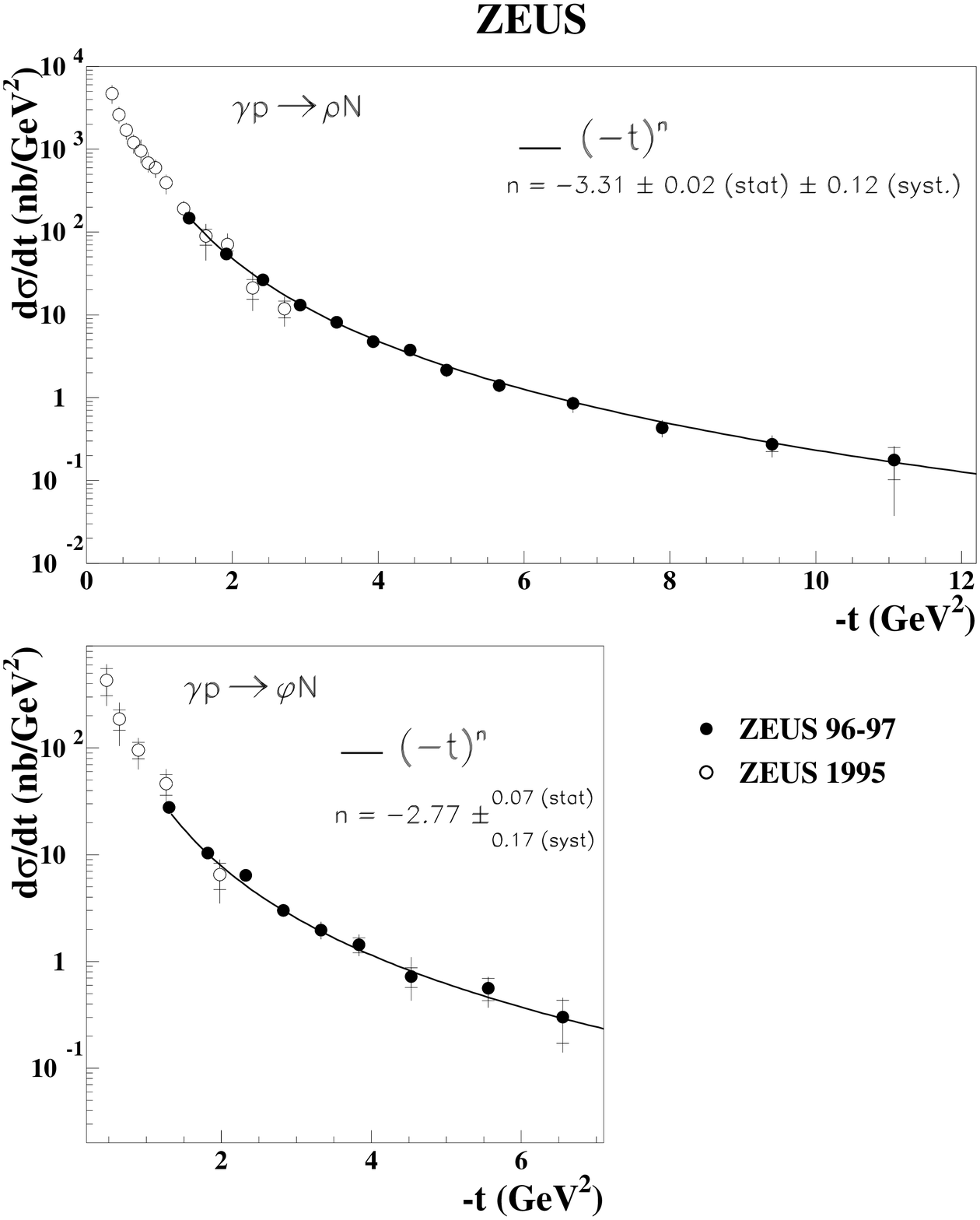,height=9cm}
~~~\epsfig{file=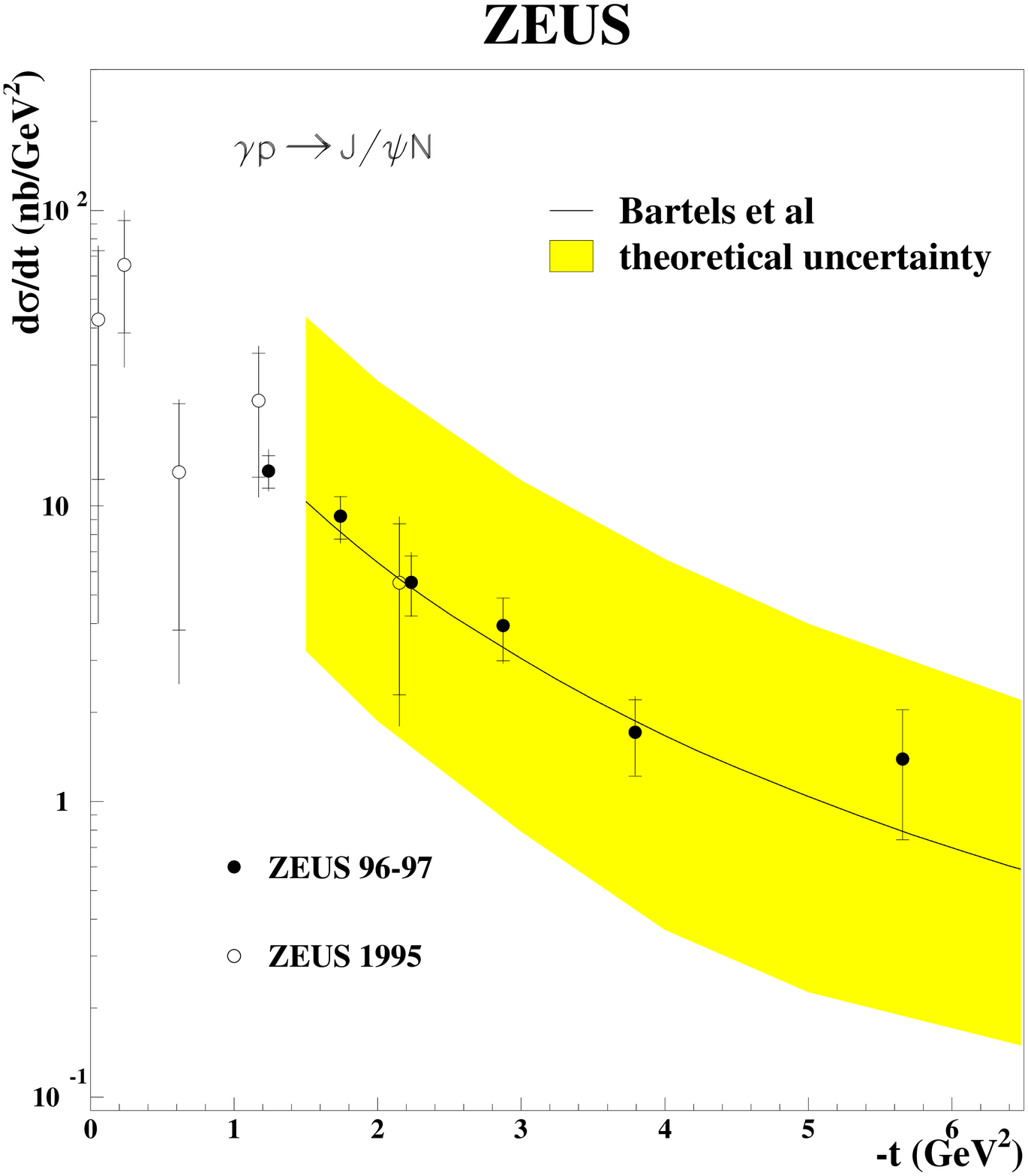,height=6.8cm}
\end{center}
\caption{ Left up - differential cross section for $\gamma^*p \rightarrow
\rho~N$, ~left down - $\gamma^*p \rightarrow \phi~N$, solid lines represent
fits with formula $ \sim |t|^n$. Right - $\gamma^*p \rightarrow J/\psi~N$,
solid line represents prediction of the model described in text.
\label{fig:sigma}} 
\vspace*{-0.1cm}
\end{figure}

In principle  large  $|t|$ should provide hard QCD scale however this cannot be 
tested in elastic VM production  where typical values of $|t|$ are lower than
$1 ~GeV^2$. That is why the ZEUS collaboration preformed a dedicated
measurement~\cite{lt} of proton-dissociative photoproduction of vector mesons
in the energy range $ 80 < W < 120~GeV$ reaching the value of $|t| \simeq 12
~GeV^2 $. In this process, $\gamma^*p \rightarrow V~N$, vector meson and
system $N$ of  proton dissociation products are separated by large rapidity
gap so mechanism of Pomeron (or colour singlet) exchange can be invoked.
The differential cross sections $d\sigma/dt$ are presented in
Figure~\ref{fig:sigma}. Contrary to elastic VM production they do not follow
the exponential  but rather power law dependence $ \sim |t|^n$, natural in pQCD
framework. The heavier VM the less steep t dependence of the cross section: 
$n = -3.31\pm0.02(stat.)\pm0.12(syst.)$ for $\rho$, 
~$n =-2.77\pm0.07(stat.)\pm0.17(syst.)$ for $\phi$ and 
~$n =-1.7\pm0.2(stat.)\pm0.2(syst.)$ for $J/\psi$ meson.
In case of $ J/\psi $ production the model~\cite{ba} assuming pQCD ``Pomeron''
coupling to single parton in the proton describes the data. 

The ratios of cross sections $\phi/\rho $ and $\psi/\rho$ 
are the next subject of interest as their values reflect the underlying
mechanism of VM production. In particular if photon couples directly to quarks
in VM and assuming flavour independent production mechanism
\begin{figure}[!] 
\vspace*{-0.2cm}
\begin{center} 
\epsfig{file=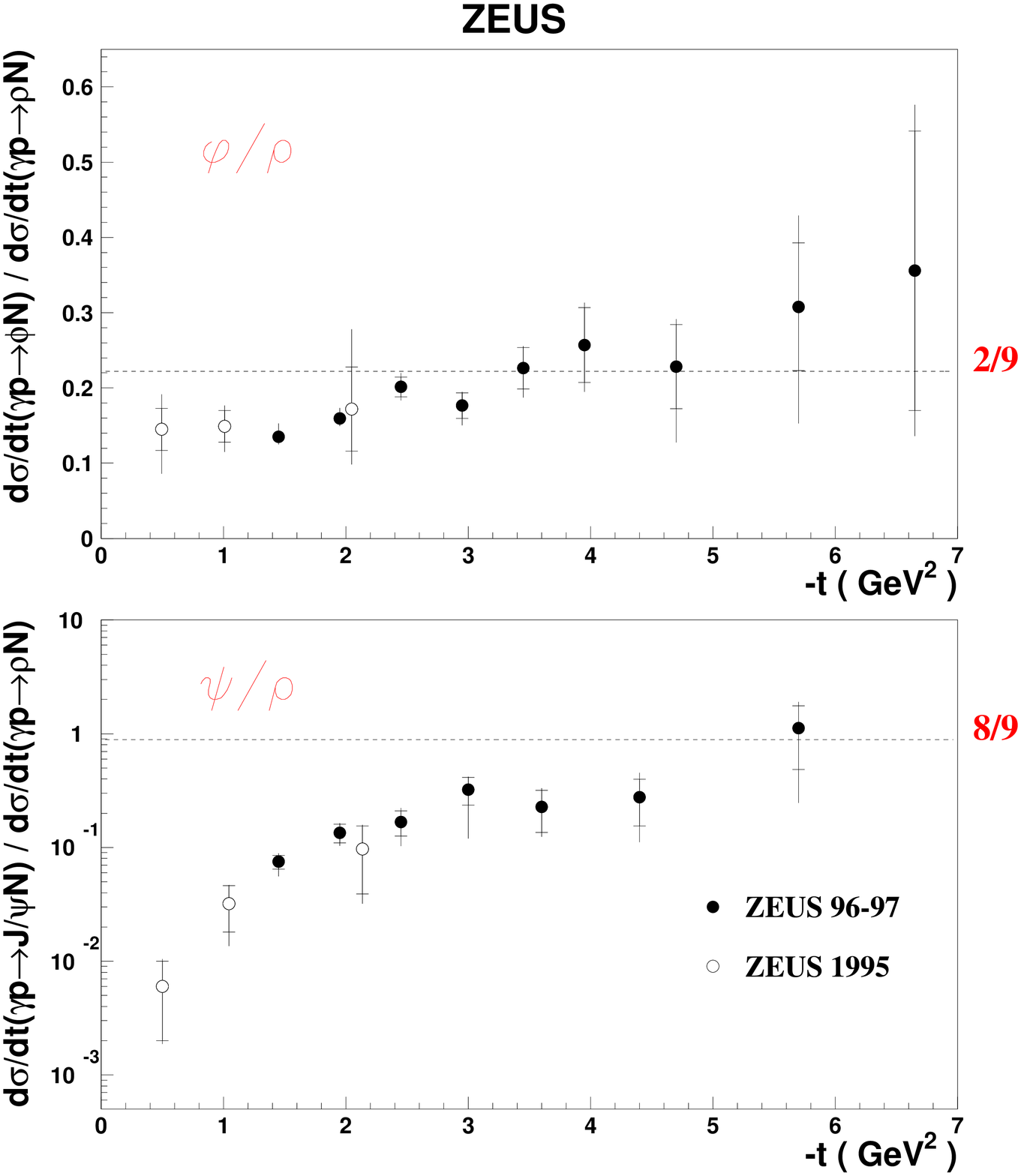,height=8.9cm}
~~~~~\epsfig{file=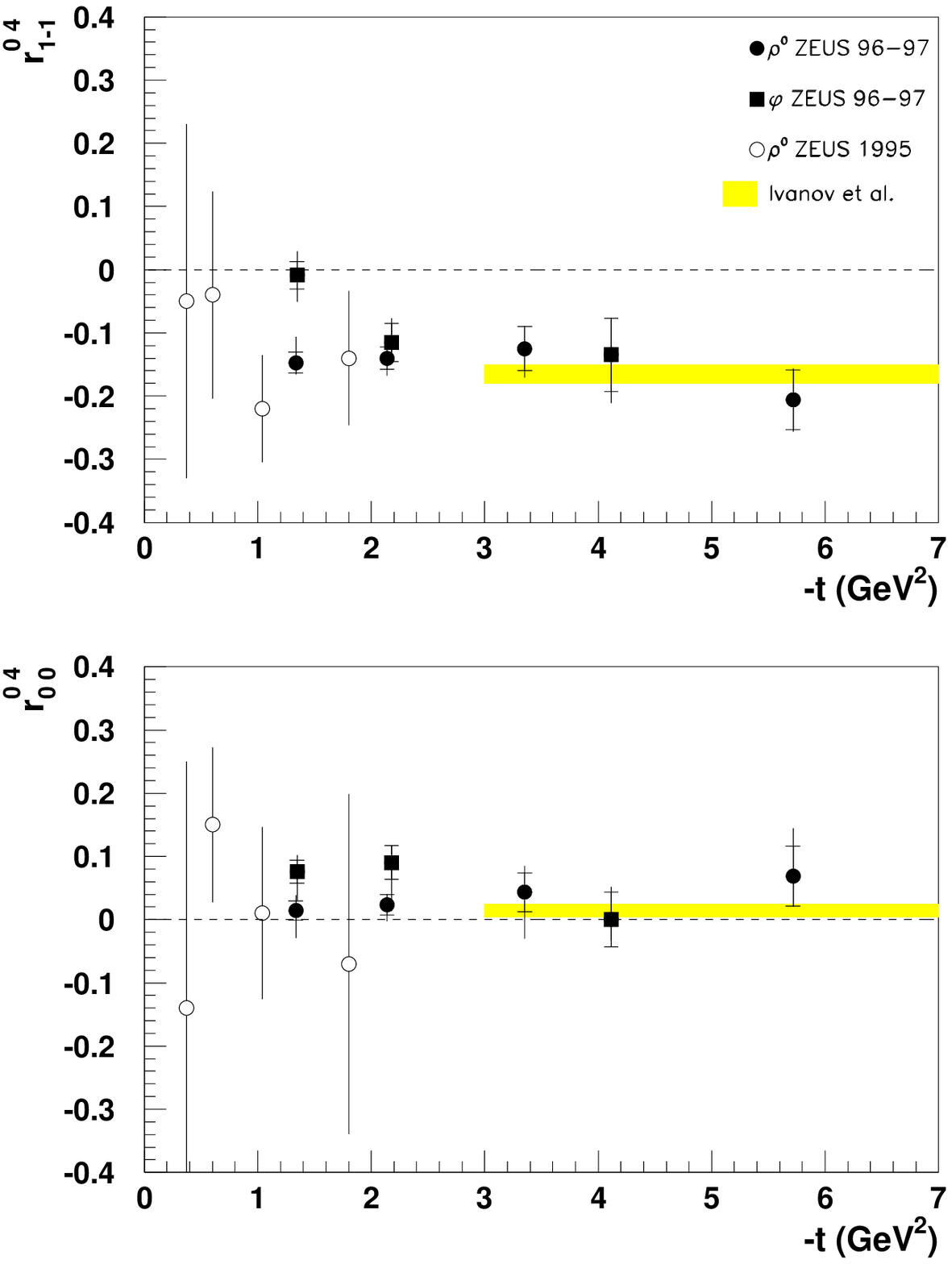,height=9.6cm}
\end{center}
\caption{ Left - the cross sections ratios $\phi/\rho $ and $\psi/\rho$ in
function of $|t|$ , ~~right - the spin density matrix elements for $\rho$ and
$\phi$ mesons in function of $|t|$ , the grey bands represent predictions
 of the pQCD based model.
\label{fig:ratios-heli}} 
\end{figure}
which is natural in ``hard'' interactions, these ratios are determined by
quark composition of  vector mesons: ~~$\phi : \psi : \rho$ = ~~~2 : 8 : 9. 
These asymptotic values of the ratios were already observed at larger  $Q^2$
values~\cite{psiz,om}. The first measurements of the VM cross section ratios
as a function of $|t|$  are presented in   Figure~\ref{fig:ratios-heli}
(left). Both ratios approach asymptotic values however the $\psi/\rho$ ratio
at higher value of $|t|$ than $\phi/\rho $ . This general behaviour is
consistent with assumption that  $|t|$ provides hard scale. 

From angular distributions (in s-channel frame) of VM decay products one can
determine its spin state which should be the same as photon if helicity is
conserved. The s-channel helicity conservation  is natural feature of
soft hadronic interactions however its small violation in  electroproduction 
of $\rho$ mesons was predicted in pQCD and observed experimentally~\cite{hel}. 
In photoproduction the nearly real photon is transversely polarised and the
same should be polarisation of VM if SCHC holds. In particular the spin density
matrix elements $r^{04}_{1-1}$ and $r^{04}_{00}$, reflecting  double and single
helicity flip    amplitudes respectively, should be zero. These matrix
elements are shown in Figure~\ref{fig:ratios-heli} (right) in function of
$|t|$ . There is clear indication of non-zero double spin flip contribution at
$|t| > 3~GeV^2$  which is reproduced by pQCD model~\cite{iv}. 
\section*{Summary}
During last year quite a lot of data on vector meson production at HERA were
released. Their statistics and quality enabled multi-variable analysis
yielding ineresting results. In particular the production of heavy mesons
shows expected features of a short distance (hard) process. For light mesons
the virtuality of the photon controls the transition between soft ($Q^2 \simeq
0$ ) and hard (``large'' $Q^2$ ) limits of strong interactions. There is
growing experimental evidence that also $|t|$ provides hard scale. Production
of vector mesons at HERA turns out to be a playground for non-perturbative and
perturbative QCD.
\section*{Acknowledgments}
I thank the DESY Directorate for their support which enabled my participation
in this very interesting conference. My work was also supported  by the Polish
State Committee for Scientific Research (KBN), grant No. 2P03B04616.

\end{document}